# Distributed Space-Time Codes for Cooperative Networks with Partial CSI

G. Susinder Rajan and B. Sundar Rajan
Department of Electrical Communication Engineering
Indian Institute of Science, Bangalore, India
Email: {susinder, bsrajan}@ece.iisc.ernet.in

***Abstract*— Design criteria and full-diversity Distributed Space Time Codes (DSTCs) for the two phase transmission based cooperative diversity protocol of Jing-Hassibi and the Generalized Nonorthogonal Amplify and Forward (GNAF) protocol are reported, when the relay nodes are assumed to have knowledge of the phase component of the source to relay channel gains. It is shown that this under this partial channel state information (CSI), several well known space time codes for the colocated MIMO (Multiple Input Multiple Output) channel become amenable for use as DSTCs. In particular, the well known complex orthogonal designs, generalized coordinate interleaved orthogonal designs (GCIODs) and unitary weight single symbol decodable (UW-SSD) codes are shown to satisfy the required design constraints for DSTCs. Exploiting the relaxed code design constraints, we propose DSTCs obtained from Clifford Algebras which have low ML decoding complexity.**

## I. Introduction & Background

Cooperative communication for wireless networks has received much attention in recent years since this technique enables users to achieve spatial diversity even if a local antenna array is not available. The diversity so obtained due to cooperation is called 'cooperative diversity'. A cooperative diversity protocol is one which dictates how users would cooperate among themselves to achieve the required diversity order. Several cooperative diversity protocols have been proposed in the literature [1]-[12]. In this paper, we focus on the two phase based cooperative diversity protocols of Jing-Hassibi [8] and the GNAF (Generalized Non-orthogonal Amplify and Forward) protocol of Rajan-Rajan [1] for three reasons- first, the operations at the relay nodes are considerably simplified, second, we can avoid imposing bottlenecks on the rate by not requiring the relay nodes to decode and third, the framework of distributed space-time codes (DSTCs) allows for more flexibility and higher spectral efficiency [1], [5], [8]. Every cooperative diversity protocol puts some constraints on the structure of the code. The constraints imposed by the protocols of Jing-Hassibi and the GNAF protocol do not permit the usage of arbitrary space-time codes as distributed space-time codes in the wireless network setting.

In this paper, we consider a scenario wherein the relays have knowledge of the phase component of the source to relay channel gains. This assumption is reasonable since the relays can potentially estimate their respective channel gains from the source to themselves. We refer to this kind of channel as the partial channel state information (CSI) relay channel. Under this assumption, we modify the GNAF protocol so as to essentially exploit this information for the sake of relaxing DSTC design constraints. Transmission in this protocol comprises of two phases- broadcast phase and cooperation phase. In the broadcast phase, the source broadcasts its information to the relays and the destination. In the cooperation phase, the source and the relays together employ a linear dispersion space-time code in a distributed fashion and transmit simultaneously (symbol synchronization assumed). The contributions of this paper are summarized as follows.

- DSTC design constraints are addressed for the scenario when the relays have knowledge of the phase component of the source to relay channel gains. Under this assumption it is shown that the constraints on the code structure are considerably reduced.
- Except for the Alamouti code, all other square complex orthogonal designs were previously not applicable as DSTCs [8]. The new relaxed constraints admit the use of any square complex orthogonal design as DSTC in the partial CSI relay channel.
- Exploiting the relaxed constraints we propose a class of four group ML decodable codes using the recently constructed Clifford Unitary Weight Single Symbol decodable (CUW-SSD) codes [19].
- Moreover it is shown that knowledge of partial CSI at the relays does not improve the optimal D-MG (Diversity Multiplexing gain) tradeoff of the protocol.

The rest of the paper is organized as follows. In Section II, we briefly describe the modifications to the GNAF protocol and state the relaxed design constraints. Also we derive a design criteria for achieving full diversity in the partial CSI relay channel based on pair-wise error probability (PEP) analysis and propose several DSTCs which exploit the relaxed design constraints. In Section III, we show that the optimal DM-G tradeoff of the protocol does not improve even if partial CSI is available at the relays. Discussions on further work comprise Section IV.

**Notation:** For a complex matrix $A$, $A^*$, $A^T$ and $A^H$ denote the conjugate, transpose and conjugate transpose respectively. $A_I$ denotes the real matrix obtained by taking the real parts of all the entries of the matrix $A$ and $A_Q$ denotes the real matrix obtained by taking the imaginary parts of all the entries of the matrix $A$. For a square matrix $B$, $|B|$ and $\mathrm{Tr}\,(B)$ denote the determinant and trace of the matrix $B$ respectively.

## II. Design of DSTCs for the partial CSI relay channel

In this section, we briefly describe the transmission in the modified GNAF protocol assuming partial CSI is available at the relays. Consider a wireless relay network consisting of a source node, a destination node and $R$ other relay nodes. All the nodes in the network are assumed to be equipped with a single antenna, synchronized at the symbol level and are subject to a half-duplex constraint, i.e., a node cannot transmit and receive simultaneously. The channel path gains from the $j^{th}$ relay to the destination denoted by $g_j$ are assumed to be i.i.d $\mathcal{CN}(0,1)$. The channel path gain, $g_0$ from the source to the destination is also assumed to be $\mathcal{CN}(0,1)$. Since the relays know the phase component of the source to the relay channel gains, we assume that the channel path gains from the source to the $i^{th}$ relay, denoted by $f_i$ are i.i.d Rayleigh random variables with mean 0 and variance 1. The signal model for the modified GNAF protocol is as shown in (1)

$$\begin{aligned} y_{D,1} &= \sqrt{\pi_1 P} g_0 s + w_1 \\ r_i &= \sqrt{\pi_1 P} f_i s + v_i, \quad i = 1,\dots,R \\ t_i &= \sqrt{\tfrac{\pi_3 P}{\pi_1 P+1}} \left( A_i r_i + B_i r_i^* \right), \quad i = 1,\dots,R \\ y_{D,2} &= \sqrt{\tfrac{\pi_3 P}{\pi_1 P+1}} \textstyle\sum_{i=1}^{R} g_i \left( A_i r_i + B_i r_i^* \right) \\ &\quad + \sqrt{\pi_2 P} g_0 \left( A_0 s + B_0 s^* \right) \\ &\quad + \sqrt{\tfrac{\pi_3 P}{\pi_1 P+1}} \textstyle\sum_{i=1}^{R} g_i ( A_i v_i + B_i v_i^* ) + w_2 \end{aligned} \tag{1}$$

where

- $s$ is the vector transmitted by the source from a codebook consisting of $\mathscr{C} = \{s_1, s_2, \dots, s_L\}$ complex vectors of size $T_1 \times 1$ satisfying $\mathrm{E}\left\{ s^H s \right\} = 1$
- $y_{D,1}$ denotes the received vector at the destination in the broadcast phase,
- $r_i$ denotes the received vector at the $i^{th}$ relay
- $t_i$ denotes the vector transmitted by the $i^{th}$ relay
- $A_i$ and $B_i$ are complex matrices of size $T_2 \times T_1$ satisfying $\| A_i \|_F^2 + \| B_i \|_F^2 \leq 1$. The pair of matrices $A_i$ and $B_i$ will be called the 'relay matrix pair' for the $i^{th}$ relay
- The quantities $\pi_1, \pi_2$ and $\pi_3$ are the power allocation factors satisfying $\pi_1 + \pi_2 + R\pi_3 = T_1 + T_2$ so that $P$ represents the total average power spent by the source and the relays together
- $y_{D,2}$ is the received vector at the destination in the cooperation phase.

The received vector at the destination can be written as

$$y = \begin{bmatrix} y_{D,1} \\ y_{D,2} \end{bmatrix} = \sqrt{\frac{\pi_3 \pi_1 P^2}{\pi_1 P + 1}} S H + W$$

where $S$ is given by (2) shown at the top of the next page, and

$$\begin{aligned} H^T &= \begin{bmatrix} g_0 & g_1 f_1 & \dots & g_R f_R \end{bmatrix} \\ W &= \begin{bmatrix} w_1 \\ \sqrt{\frac{\pi_3 P}{\pi_1 P+1}} \sum_{i=1}^{R} g_i \left( A_i v_i + B_i v_i^* \right) + w_2 \end{bmatrix}. \end{aligned}$$

We refer to the collection of all the $(T_1 + T_2) \times (R+1)$ matrices $S$ as the DSTC. To make the PEP analysis simpler, we will require the noise components of $n_i = A_i v_i + B_i v_i^*$ to be uncorrelated. For this to happen the matrix

$$Z_i = \begin{bmatrix} A_{iI} + B_{iI} & -A_{iQ} + B_{iQ} \\ A_{iQ} + B_{iQ} & A_{iI} - B_{iI} \end{bmatrix}$$

should satisfy the condition that $Z_i Z_i^T$ is a diagonal matrix $\forall\ i = 1,\dots,R$. The following theorem states that the well known rank and determinant criteria continue to hold even for the partial CSI relay channel.

*Theorem 1:* If the DSTC $S$ is such that the matrix $Z_i = \begin{bmatrix} A_{iI} + B_{iI} & -A_{iQ} + B_{iQ} \\ A_{iQ} + B_{iQ} & A_{iI} - B_{iI} \end{bmatrix}$ satisfies the condition $Z_i Z_i^T$ is a diagonal matrix $\forall\ i = 1,\dots,R$ and if $\Delta S = S_i - S_j$ has full rank for all pairs of distinct codewords $S_i$ and $S_j$, then full diversity of order $R+1$ will be achieved when a ML receiver is employed at the destination. Further for large $P$, the coding gain is controlled by the minimum determinant over all possible codeword difference matrices $\Delta S$.

*Proof:* The proof follows the same proof techniques as in [1], [8] and hence omitted. ∎

Observe that each column of $S$ can have both the variables as well as their conjugates. Note that this was forbidden by the constraints imposed by the protocol of [8] wherein either the variables or their conjugates were allowed in any column of $S$. Though the GNAF protocol of [1] allowed both variables as well as their conjugates, PEP analysis showed that the diversity depended on the rank of the difference of some other matrix manufactured from $S$ but not on $S$ itself, thus complicating code design further. But here we see that *if the relays have partial CSI, then the DSTC design problem becomes very similar to the code design problem for colocated MIMO communication except for the additional constraint on the matrix* $Z_i$. From the structure of $S$ in (2), we see that for all code construction purposes we can consider only the submatrix given by

$$\begin{bmatrix} A_1 s + B_1 s^* & A_2 s + B_2 s^* & \cdots & A_R s + B_R s^* \end{bmatrix}$$

since if the STBC corresponding to this sub-matrix is fully diverse, then the DSTC $S$ is also fully diverse. Hence throughout the rest of the paper we consider designing only this submatrix.

The following lemma can be used as a sufficient condition to check whether a STBC respects the constraints required for the partial CSI relay channel.

*Lemma 1:* Any linear dispersion code $\mathcal{C}$ in $T_1$ complex variables having the following two properties satisfies the constraints on the matrix $Z_i$

- The real and imaginary part of any entry of $\mathcal{C}$ should be one of the terms $\{\pm\alpha_1 x_{1I}, \dots, \pm\alpha_{T_1} x_{T_1 I}, \pm\beta_1 x_{1Q}, \dots, \pm\beta_{T_1} x_{T_1 Q}, 0\}$
- In any column of $\mathcal{C}$, any of the terms $\{\pm\alpha_1 x_{1I}, \dots, \pm\alpha_{T_1} x_{T_1 I}, \pm\beta_a x_{1Q}, \dots, \pm\beta_{T_1} x_{T_1 Q}\}$ can occur at most once, where $\alpha_i$'s and $\beta_i$'s are real scalars.

$$S = \begin{bmatrix} \sqrt{\frac{\pi_1 P+1}{\pi_3 P}} I_{T_1} s & 0 & \dots & 0 \\ \sqrt{\frac{\pi_2(\pi_1 P+1)}{\pi_3 \pi_1 P}} A_0 s + B_0 s^* & A_1 s + B_1 s^* & \dots & A_R s + B_R s^* \end{bmatrix} \quad (2)$$

*Proof:* The matrix $Z_i$ is precisely the matrix that linearly transforms the vector $\begin{bmatrix} s_I \\ s_Q \end{bmatrix}$ to generate the stacked column vector $\begin{bmatrix} G_{iI} \\ G_{iQ} \end{bmatrix}$ consisting of the real and imaginary parts of the $i^{th}$ column $G_i$ of $\mathcal{C}$, i.e.,

$$\begin{bmatrix} G_{iI} \\ G_{iQ} \end{bmatrix} = Z_i \begin{bmatrix} s_I \\ s_Q \end{bmatrix}.$$

The first property ensures that every row of the matrix $Z_i$ has at the most only one nonzero entry. The second property in addition to the first property ensures that different rows have the nonzero entries necessarily in different columns and hence the rows of $Z_i$ are orthogonal. Thus $Z_i Z_i^T$ is a diagonal matrix. ■

Proposition 1 shows that several known DSTCs provide full diversity for the partial CSI relay channel. We briefly introduce these codes before proving the proposition.

*Definition 1:* A Generalized Co-ordinate Interleaved Orthogonal Design (GCIOD) [15] of size $N_1 \times N_2$ in variables $x_i, i = 0, \cdots, K-1$ (where $K$ is even) is a $L \times N$ matrix $S$, such that

$$S(x_0, \cdots, x_{K-1}) = \begin{bmatrix} \Theta_1(\tilde{x}_0, \cdots, \tilde{x}_{K/2-1}) & 0 \\ 0 & \Theta_2(\tilde{x}_{K/2}, \cdots, \tilde{x}_{K-1}) \end{bmatrix}$$

where $\Theta_1(x_0, \cdots, x_{K/2-1})$ and $\Theta_2(x_{K/2}, \cdots, x_{K-1})$ are Complex Orthogonal Designs (CODs) of size $L_1 \times N_1$ and $L_2 \times N_2$ respectively, with rates $K/2L_1, K/2L_2$ respectively, where $N_1 + N_2 = N$, $L_1 + L_2 = L$, $\tilde{x}_i = \Re\{x_i\} + \jmath\Im\{x_{(i+K/2)_K}\}$ and $(a)_K$ denotes $a \pmod K$. If $\Theta_1 = \Theta_2$ then we call this design a *Co-ordinate interleaved orthogonal design(CIOD)*

In this paper, we follow the definition of complex orthogonal designs as given in [20].
Let

$$S_U = \sum_{i=1}^{K} (x_{iI} C'_{iI} + x_{iQ} C'_{iQ}). \quad (3)$$

be a Unitary Weight code (UW STBC), i.e., for which all the weight matrices $C'_{iI}$ and $C'_{iQ}$ are unitary. We normalize the weight matrices of the code as

$$\begin{aligned} C_{iI} &= C'^H_{1I} C'_{iI} \\ C_{iQ} &= C'^H_{1I} C'_{iQ} \end{aligned}$$

to get *the normalized version* of (3) to be

$$S_N = x_{1I} I_n + x_{1Q} C_{1Q} + \sum_{i=2}^{K} (x_{iI} C_{iI} + x_{iQ} C_{iQ}). \quad (4)$$

and call the code $S_N$ to be the normalized code of $S_U$.

A $n \times n$ UW code described by (3) is called a Clifford Unitary Weight SSD (CUW-SSD) code if after normalization the weight matrices satisfy the following conditions [19]:

$$\begin{aligned} C_{iI}^H &= -C_{iI} \quad 2 \le i \le K \\ C_{iI} C_{jI} &= -C_{jI} C_{iI}, \quad 2 \le i \ne j \le K \\ C_{1Q}^H &= C_{1Q} \\ C_{iQ} &= C_{1Q} C_{iI}, \quad 2 \le i \le K \\ C_{1Q} C_{jI} &= C_{jI} C_{1Q} \quad 1 \le j \le K \end{aligned} \quad (5)$$

The name in the above definition is due to the fact that such codes are constructable using matrix representations of real Clifford algebras.

*Proposition 1:* The following STBCs can be used on the partial CSI-relay channel and also provide full diversity for the system.

- Complex Orthogonal designs
- Clifford Unitary Weight-Single Symbol Decodable (CUW-SSD) [19] STBCs as constructed in [19]
- Generalized Co-ordinate Interleaved Orthogonal Designs (GCIOD) [15]

*Proof:* We shall prove the proposition by showing that the two conditions stated in Lemma 1 are satisfied by these codes. Complex orthogonal designs satisfy the required conditions stated in Lemma 1 and these have already been stated in [20]. As a consequence, it is immediate from the construction of GCIOD [15] that GCIOD also satisfy the required conditions. Let the weight matrices of the CUW-SSD code constructed in [19] be $I, C_{2I}, \dots, C_{T_1 I}, C_{1Q}, \dots, C_{T_1 Q}$. Since the weight matrices are unitary and have the property that its entries are only $0, \pm 1, \pm j$, any column of any weight matrix contains exactly one non-zero entry. Thus the second condition of Lemma 1 is satisfied. Also note that

$$\begin{aligned} C_{iI}^H C_{iQ} + C_{iQ}^H C_{iI} &= C_{iI}^H C_{1Q} C_{iI} + C_{iI}^H C_{1Q}^H C_{1I} \\ &= 2C_{1Q} \end{aligned}$$

where, the second equality follows from the fact that $C_{1Q}$ is a Hermitian matrix which commutes with all the other weight matrices. If $A$ and $B$ are two square matrices of the same size, we shall refer to the matrix $A^H B + B^H A$ as the outer product between them. By virtue of the construction in [19], the matrix $C_{1Q}$ has all its diagonal entries to be equal to zero. Therefore the diagonal entries of the outer product between any two weight matrices is zero. We use this property to prove by contradiction that the first condition in Lemma 1 is satisfied. Supposing the first condition of Lemma 1 is not satisfied. Then the real part or imaginary part of some entry in the linear design should contain more than one of the terms $\{x_{1I}, \dots, x_{T_1 I}, x_{1Q}, \dots, x_{T_1 Q}\}$. Let the position of that entry be denoted by $(i, j)$. Now consider the $j^{th}$ columns of the weight matrices of the corresponding terms in the $(i, j)$ entry. Now these column vectors will have non-zero entries in the same position. Now take any two such column vectors. We will show that the non-zero entry in both the columns cannot be both real or both imaginary. This will contradict our assumption that the first condition of Lemma 1 is not satisfied.

Now supposing the entries were both real, then it will contradict the fact that the outer product between their corresponding weight matrices has zero diagonal entries. Similarly, if they were both imaginary also it leads to the same contradiction. Therefore if one column has non-zero real entry, the other column has to necessarily have non-zero imaginary entry. This proves the proposition. ■

### *A. Low ML decoding complexity DSTCs for partial CSI relay channel*

In this subsection, we explicitly construct a class of $4$-group ML decodable DSTCs for the partial CSI relay channel. The ML decoding of a DSTC in $K$ real variables $x_1, \ldots, x_K$ is in general, joint decoding of all the $K$ variables. However if $K = g\lambda$ and if the variables can be partitioned into $g$ subsets each containing $\lambda$ number of variables such that the ML decoding can be done for a subset independently of the variables of other subsets, then the code is said to be $g$-group ML decodable.

Consider the following example of a design for $4$ relays.

*Example 1:* The $4\times 4$ design $C_{CA}$ shown below for $4$ relays was obtained using Clifford Algebras [19].

$$\begin{bmatrix} x_{1I} - ix_{4Q} & x_{2I} + ix_{3I} & x_{4I} + ix_{1Q} & -x_{3Q} + ix_{2Q} \\ -x_{2I} + ix_{3I} & x_{1I} + ix_{4Q} & -x_{3Q} - ix_{2Q} & -x_{4I} + ix_{1Q} \\ -x_{4I} - ix_{1Q} & x_{3Q} - ix_{2Q} & x_{1I} - ix_{4Q} & x_{2I} + ix_{3I} \\ x_{3Q} + ix_{2Q} & x_{4I} - ix_{1Q} & -x_{2I} + ix_{3I} & x_{1I} + ix_{4Q} \end{bmatrix}$$

The design $C_{CA}$ is known to be a single-symbol decodable code which achieves full diversity in the co-located MIMO channel for an appropriately chosen signal set. More importantly this design has the property that $Z_i Z_i^T = I_8, i = 1, \ldots, 4$. This results in the covariance matrix of the noise vector

$$\left[ \sqrt{\frac{\pi_3 P}{\pi_1 P + 1}} \sum_{i=1}^{R} g_i \left( A_i v_i + B_i v_i^* \right) + w_2 \right]$$

to be a scaled identity matrix. Let $\Omega$ be the covariance matrix of $W$. We need to premultiply $Y$ by $\Omega^{-\frac{1}{2}}$ in order to make the covariance matrix of $W$ to be identity matrix (this process is called whitening) during the process of ML decoding. This may in general disturb the low ML decodability property of DSTCs. However the above code retains the same low ML decoding complexity because $Z_i Z_i^T = I_8, i = 1, \ldots, 4$. Note that this code has *unitary weight matrices* whereas the precoded coordinate interleaved orthogonal designs (PCIOD) in [2] have *non-unitary weight matrices*. Codes with unitary weight matrices distribute the power uniformly among the relays as well as in time.

We now generalize the above DSTC for more number of relays. When the number of relays is a multiple of four, say $R = 4k$, we can put $k$ blocks of the design $C_{CA}$ along the diagonal, where each block is the same design $C_{CA}$ but in different variables. Joint precoding of one real variable from each block can be done to achieve full diversity. We now show an example construction for $R = 8$.

*Example 2:* The design for $8$ relays is given by (6), where the real variables take values as follows. The vectors $\begin{bmatrix} x_{iI} \\ x_{iQ} \\ x_{(i+4)I} \\ x_{(i+4)Q} \end{bmatrix}, i = 1, \ldots, 4$ are allowed to take values from an appropriately rotated lattice constellation to ensure full diversity. Because of the block diagonal structure of this code, $4$-group ML decodability is retained even after the process of whitening. Note that the way we choose the set of real variables to be jointly precoded is crucial. Arbitrarily choosing these set of real variables can potentially disturb $4$-group decodability although we may still get full diversity [1].

The the way we have chosen these set of real variables in the above example can be easily extended to more number of relays also. For other even number of relays, we can put a combination of PCIOD and the $4\times 4$ code in the above example on the blocks along the diagonal and precode appropriately to get full diversity and still retaining $4$-group ML decodability.

## III. D-MG TRADEOFF

The D-MG tradeoff is an approximate characterization of the tradeoff between diversity gain and multiplexing gain in the high SNR regime [17], [18]. In this section, it is shown that knowledge of partial CSI at the relays does not improve the optimal D-MG tradeoff of the protocol. Towards that end, let us first study the optimal DM-G tradeoff of the channel faced by these DSTCs. Let

$$H_1 = \begin{bmatrix} g_0 \\ g_1 f_1 \\ \vdots \\ g_R f_R \end{bmatrix} \quad \text{and} H_2 = \begin{bmatrix} g_0 \\ g_1 f_1' \\ \vdots \\ g_R f_R' \end{bmatrix}$$

where, the $g_i$s and $f_i$s are i.i.d complex Gaussian random variables whereas the $f_i'$s are i.i.d Rayleigh random variables. $H_1$ is the channel faced by the DSTC when partial CSI is not available at the relays. This channel was named as the 'two product' channel in [10], [11] and its optimal DM-G tradeoff was also studied. $H_2$ is the channel faced by the DSTC when partial CSI is available at the relays. We refer to this channel as the 'two product partial CSI relay channel'.

*Proposition 2:* The optimal D-MG tradeoff of the two product partial CSI relay channel is same as that of the two product channel.

*Proof:* The optimal D-MG tradeoff of the two product partial CSI relay channel depends upon the quantity $|I + \rho H_2^H H_2|$ where $\rho$ is the SNR. But this is equal to $1 + \rho \sum_{i=0}^{R} |f_i'|^2 |g_i|^2$. Similarly

$$|I + \rho H_1^H H_1| = 1 + \rho \sum_{i=0}^{R} |f_i|^2 |g_i|^2.$$

But the resultant distributions are same for both the two product partial CSI relay channel as well as the two product channel. This completes the proof. ■

[1]For example, joint precoding of all the real variables will make the code 1-group ML decodable.

$$\left[\begin{array}{cc} \begin{array}{cccc} x_{1I}-ix_{4Q} & x_{2I}+ix_{3I} & x_{4I}+ix_{1Q} & -x_{3Q}+ix_{2Q} \\ -x_{2I}+ix_{3I} & x_{1I}+ix_{4Q} & -x_{3Q}-ix_{2Q} & -x_{4I}+ix_{1Q} \\ -x_{4I}-ix_{1Q} & x_{3Q}-ix_{2Q} & x_{1I}-ix_{4Q} & x_{2I}+ix_{3I} \\ x_{3Q}+ix_{2Q} & x_{4I}-ix_{1Q} & -x_{2I}+ix_{3I} & x_{1I}+ix_{4Q} \end{array} & \mathbf{0} \\ \mathbf{0} & \begin{array}{cccc} x_{5I}-ix_{8Q} & x_{6I}+ix_{7I} & x_{8I}+ix_{5Q} & -x_{7Q}+ix_{6Q} \\ -x_{6I}+ix_{7I} & x_{5I}+ix_{8Q} & -x_{7Q}-ix_{6Q} & -x_{8I}+ix_{5Q} \\ -x_{8I}-ix_{5Q} & x_{7Q}-ix_{6Q} & x_{5I}-ix_{8Q} & x_{6I}+ix_{7I} \\ x_{7Q}+ix_{6Q} & x_{8I}-ix_{5Q} & -x_{6I}+ix_{7I} & x_{5I}+ix_{8Q} \end{array} \end{array}\right] \qquad (6)$$

Thus from the above proposition and the results in [10], [11] it follows that the optimal D-MG tradeoff of the two product partial CSI relay channel is given by

$$d^*(r) = (R+1)(1-r)^+.$$

which is the same as that of the colocated MISO (Multiple Input Single Output) channel.

## IV. Discussions

We have thus shown that having partial CSI at the relays helps in relaxing the constraints on the structure of the DSTC. In particular, we have show that the conjugate linearity property (any column of the design should have only the variables or only their conjugates) [1] need not be satisfied. We have also proposed few DSTCs exploiting these relaxed conditions. These relaxed conditions might help in constructing DSTCs with low decoding complexity or with more coding gain. Recently in [21], the problem of constructing Distributed Orthogonal Space-Time codes was addressed and the authors derived bounds on the maximal rate of such codes. These Distributed Orthogonal Space-Time Codes (DOSTCs) are nothing but a special class of Orthogonal Designs that satisfy the additional requirements of the protocol. It would be interesting to know whether the relaxed conditions for the partial CSI relay channel aids in constructing DOSTCs with higher rates.

## Acknowledgment

This work was partly supported by the DRDO-IISc Program on Advanced Research in Mathematical Engineering, partly by the Council of Scientific & Industrial Research (CSIR), India, through Research Grant (22(0365)/04/EMR-II) to B.S. Rajan. The authors are grateful to the anonymous reviewers for their valuable comments which helped improve the clarity of the paper.